\documentclass[letterpaper]{article} 
\usepackage{aaai2027}  
\usepackage[hyphens]{url}  
\usepackage{graphicx} 
\usepackage{natbib}  
\usepackage{caption} 
\usepackage[utf8]{inputenc}
\usepackage{amsmath}
\usepackage{amssymb}
\usepackage{booktabs}
\usepackage{siunitx}
\usepackage{tabularx}
\usepackage{ragged2e}

\title{Who's Blocking Whom? Candidate Generation and Block Prediction on Bluesky}

\author{
Cecilia Galbiati,
Carlo Bono,
Carlo Piccardi,
Francesco Pierri
}

\affiliations{
Politecnico di Milano
}

\begin{document}

\maketitle

\begin{abstract}
Blocking is a widely used tool that helps people manage unwanted interactions on social platforms.
We study the problem of predicting block events on Bluesky: whether a given user will block a particular account, given recent interaction, network, activity, and content signals.
Using more than three million block events and over 260 million user interactions, we examine several formulations of the directed block prediction problem, differing in which possible targets are considered and how negative examples are selected.
We observe that only about 5\% of user blocks are preceded by a recent direct interaction.
Expanding the set of possible targets to include accounts connected through a common neighbor raises the share to about 22\%, once high-degree accounts are excluded as intermediaries.
Conditional on candidate inclusion, Hits@1 ranges from $61.3\%$ to $72.4\%$, compared with a random baseline of $16.7\%$.
However, predictive performance and the signals used by the models depend strongly on how candidates and comparison examples are constructed.
Although our goal is to empirically analyze user behavior rather than to propose a deployable system, these findings are relevant to user-facing moderation tools that might help users identify accounts they may wish to avoid or block. In such tools, candidate generation would be a substantive design and evaluation choice rather than merely a preprocessing step.
\end{abstract}

\section{Introduction}
Blocking is a widely used form of user-level moderation that allows
people to limit unwanted contact and shape their experience on social
platforms~\cite{vogels2021, seering2020}. Unlike unfollowing or muting,
blocking directly changes how two accounts can interact. Users may block
others in response to harassment or unwanted attention, but blocking may
also reflect political disagreement, content preferences, and community
norms~\cite{baysha2020, martel2024blocking, kaiser2022}.

Blocking behavior is difficult to study on most platforms because
moderation actions are generally private. 
Bluesky provides an unusual
empirical setting: its public protocol makes block events and other user
interactions observable at scale~\cite{kleppmann2024bluesky}.
This transparency has enabled
recent work to examine the characteristics of blocked accounts and the
role of individual and community-level blocking in the platform's
moderation ecosystem~\cite{bono2026selfmoderation, sokoto2026open}.

Existing predictive work has primarily studied blocking at the account level. \citet{bono2026selfmoderation} examine whether an account is generally likely to be blocked based on its activity, content, and network position. 
Related platform-wide work similarly compares the characteristics of blocked and non-blocked accounts \cite{sokoto2026open}. 
These studies establish that blocking outcomes are associated with observable user behavior, but they do not identify who is likely to block whom. 
We directly address this gap: given users $u$ and $v$ and their recent activity observed before day $t$, can we predict whether $u$ will block $v$ on day $t$?

Dyadic (i.e., pair-specific and directed) block prediction requires defining a manageable set of candidate targets.
Considering every possible ordered pair on the platform would be computationally impractical, so a prediction framework must restrict attention to plausible targets identified from prior observed activity.

candidate generation is therefore part of the prediction problem itself: it determines which accounts are compared and which future block events can, in principle, be captured. 
A model may rank candidates accurately when the true target is included, yet still miss most future blocks because the candidate-generation rule excludes their targets.

Using more than three million block events and over 260 million interactions observed on Bluesky, we study how candidate-selection choices affect directed block prediction. 
We evaluate several experimental formulations that vary in how candidate targets are generated, how negative examples are constructed, and whether candidates are evaluated globally or separately for each potential blocker.
We assess both candidate coverage---the proportion of observed future blocks whose true targets are admitted by a candidate generation rule---and predictive performance conditional on the true target being included in the resulting candidate set. 
We also compare multiple prediction models and examine the contributions of different groups of features.

This paper makes four contributions:
\begin{enumerate}
    \item We formulate pair-specific block prediction on Bluesky as a two-stage
    temporal task: identifying plausible block targets and then
    predicting or ranking them.

    \item We quantify how much blocking behavior can be captured under
    different candidate-generation rules. Recent direct interaction
    covers only $5.2\%$ of future blocks, while also considering accounts
    connected through one shared interaction partner raises the estimated
    coverage to approximately $22.0\%$ after excluding highly connected
    intermediaries.

    \item We show that predictive performance depends strongly on how
    candidate targets and comparison examples are selected. We evaluate
    several experimental formulations, ranging from predicting among
    recently interacting pairs to ranking multiple possible targets for
    the same user, including candidates drawn from the user's local
    interaction network.

    \item We show that experimental design also changes which signals
    appear most informative to the model. Features that are important
    in global classification may become uninformative when candidate
    targets are compared separately for each user, while signals based
    on prior blocks received raise important risks of reinforcing
    visibility or collective exclusion.
\end{enumerate}

We emphasize that our primary aim is to use prediction as a way to study blocking behavior, not to propose a system for deployment.
Nevertheless, the findings are relevant to moderation tools that might surface accounts a user may wish to avoid or block. 
Assessing such tools requires attention not only to how accurately they rank a given set of accounts, but also to how that set is constructed, which patterns of behavior the model relies on, and whether its suggestions could limit user choice or reinforce unfair exclusion. 
Similar concerns arise for other sparse, pair-specific actions, including reporting, muting, and unfollowing.
The framework could also be applied to other social platforms, such as X and Mastodon, provided that comparable temporal data on user interactions and moderation actions are available.

\section{Related Work}
\label{sec:related-work}

\paragraph{Temporal and negative-link prediction.}
Predicting a directed block can be viewed as a generalized temporal link-prediction problem: the target edge is a future directed block, while the observed network history may include multiple types of user interaction, such as follows, likes, reposts, and replies, in addition to prior blocking activity \cite{libennowell2007linkprediction}. The field has progressed from static proximity heuristics to learned representations that account for node context and time. GraphSAGE learns inductive neighborhood aggregation \cite{hamilton2017inductive}, while TGAT and TGN incorporate temporal information through time encodings or node memories \cite{xu2020inductive,rossi2020temporal}. Gradient-boosted decision trees, such as LightGBM \cite{ke2017lightgbm}, provide a computationally efficient alternative to node-embedding and graph neural network approaches when network information can be represented as structured features. Prior work has shown that tree-based models remain highly competitive with deep learning methods on many tabular prediction tasks \cite{grinsztajn2022tree}.

The evaluation protocol is equally important in this domain. Random negatives can inflate performance and obscure inductive difficulty, motivating historical and future-aware alternatives \cite{poursafaei2022towards}; this inspires our observed-future negative definition, which excludes candidates that become positive at any later observable day. Furthermore, recent benchmarks formalize evaluation with multiple negatives, Hits@$K$, and MRR \cite{huang2023temporal}, a template we follow for our grouped settings where each positive competes against $K=5$ same-source negatives. Our study extends this evaluation concern from negative selection to candidate-set construction: before ranking candidates, one must decide which ordered pairs are eligible to be ranked.

Blocks and muting also differ fundamentally from the positive ties usually predicted. Blocks have been modeled as directed negative edges in signed networks, where local structural patterns predict edge sign \cite{leskovec2010predicting}. Work on tie dissolution emphasizes temporality, reciprocity, and attention \cite{wu2020mining,kwak2012receiver}. Blocking is more deliberate and strongly directional: field experiments report that users are substantially more likely to block counter-partisan than copartisan accounts \cite{martel2024blocking}, motivating our inclusion of explicit direction-of-exposure features rather than only interaction volume.

\paragraph{Blocking as user-controlled moderation and platform governance.}
Blocking is both a safety mechanism and a form of platform governance: moderation choices shape visibility, participation, and exclusion \cite{gillespie2018custodians,klonick2018newgovernors}. In decentralized systems, part of this authority shifts toward communities and end users, making personal moderation a question of user agency as well as governance \cite{jhaver2023multilevel}.

Prior work shows that users do not view personal controls as simple substitutes for platform enforcement. Preferences for individualized moderation vary with perceived harms and attitudes toward free expression \cite{jhaver2025individualcontrol}, while personalized tools can increase control but also shift configuration and moderation labor onto users \cite{jhaver2023personalizing}. Cross-national evidence further shows that perceived harms and preferred remedies vary across contexts \cite{schoenebeck2023majority}.

Collective blocklists can support protection from harassment but may also create false positives and perceived injustice \cite{geiger2016bot,jhaver2018online}. This is directly relevant to predictive blocking: signals based on prior blocks received may reflect harmful behavior, but may also capture controversy, coordinated reporting, or group-level exclusion.

Recent work on Bluesky has examined its network structure and temporal interactions
\cite{kleppmann2024bluesky,balduf2024looking,quelle2025bluesky,jeong2024bluetempnet},
as well as platform-wide blocking behavior
\cite{sokoto2026open,bono2026selfmoderation}.
We instead predict future directed blocks between specific user pairs and examine how candidate generation affects coverage, performance, and feature reliance.

\begin{figure*}[t]
    \centering
    \includegraphics[width=\textwidth]{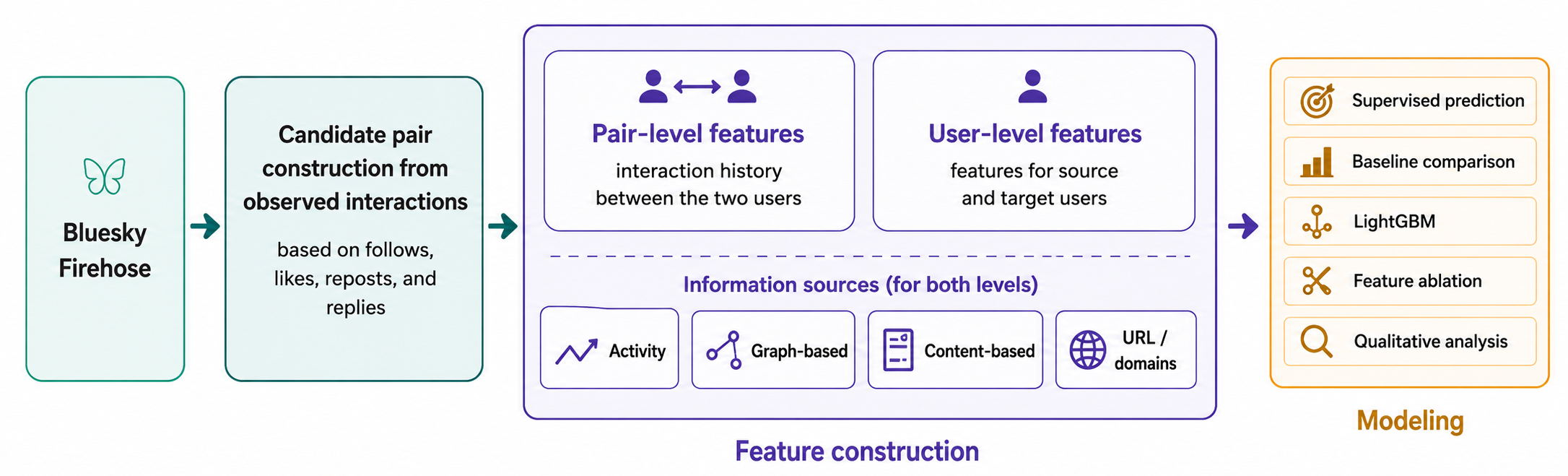}
    \caption{
        Overview of the two-stage prediction task.
        candidate generation determines which user pairs are considered.
        A supervised model then scores the admitted candidates using pair-, source-, target-, graph-, and content-level features.
        Predictive metrics are therefore conditional on the candidate set.
    }
    \label{fig:workflow}
\end{figure*}

\section{Problem Formulation and Data}
\label{sec:problem-data}

For each prediction day $t$, we consider an ordered user pair $(u,v)$, where $u$ is the potential blocker and $v$ is a candidate block target.
The ordering matters because a block initiated by $u$ toward $v$ is distinct from a block initiated by $v$ toward $u$.
We restrict the analysis to the first observed block event for each ordered pair $(u,v)$, excluding subsequent block events for pairs for which a block has already been observed\footnote{Blocks can be revoked by users.}. Since the observation period begins on June 1, this does not necessarily correspond to the pair’s first-ever block on Bluesky.
Within this filtered event set, we define $y_{u,v,t}=1$ when the first observed block from $u$ to $v$ occurs on day $t$. Negative instances are constructed according to the \emph{same-day} or \emph{observed-future} rules described below.

Features are constructed from activity observed during the \num{7} days before the prediction day $t$.
We denote this historical window by $[t-W,t-1]$, with $W=\num{7}$ (days).

We use two definitions of negative examples throughout our experiments.
A \emph{same-day negative} is a candidate pair for which no block is observed on day $t$.
An \emph{observed-future negative} must remain unblocked from day $t$ until the end of the observation period.
This more conservative definition is intended to avoid labeling as negative a pair that is subsequently observed to become positive.
Because the observation period has a fixed end date, however, the resulting follow-up horizon varies with the prediction day.
We therefore exclude the final seven prediction days, ensuring that every retained instance has at least seven days of observable future, although the follow-up duration is not uniform across instances.

For each selected candidate, we construct a feature vector $\mathbf{x}_{u,v,t}$ from activity observed during the historical window ending in $t-1$. 
The feature vector is organized into pair-, source-, target-, graph-, and content-level features, as illustrated in Figure~\ref{fig:workflow}.
Pair-level features describe whether and how the two users interacted, including interaction direction, type, frequency, recency, and the number of interaction types involved.
Source- and target-level features summarize, for $u$ and $v$ respectively, user activity, graph position, visibility, and recent blocking history, including both performed and received blocks.
Graph-level features describe the local structure of the historical interaction graph constructed from follows, likes, reposts, and replies observed during $[t-W,t-1]$.
Content-level features include post-level statistics as well as URL- and domain-sharing activity, following \citet{bono2026selfmoderation}.


We analyze Bluesky block and interaction events observed from June~\num{1} to August~\num{28}, \num{2024}.
After preprocessing, the dataset contains \num{2561333} distinct users and \num{3071079} first observed user-to-user block events.
These events involve \num{219487} blockers and \num{594131} block targets.
The dataset also contains \num{263416428} timestamped follows, likes, reposts, and replies.

Bluesky makes platform activity publicly accessible through its Firehose, including the block and interaction events examined in this study. We use the dataset collected by \citet{seckin2025rise} from this public stream. Although individual records are publicly observable, bulk user-to-user block data reveal potentially sensitive relationships between identifiable accounts. To reduce privacy risks, we do not release the raw block or interaction records.

\begin{table}[t]
\centering
\small
\setlength{\tabcolsep}{3pt}
\renewcommand{\arraystretch}{1}

\begin{tabular}{
c
>{\raggedright\arraybackslash}p{0.25\columnwidth}
>{\raggedright\arraybackslash}p{0.25\columnwidth}
>{\raggedright\arraybackslash}p{0.30\columnwidth}
}
\toprule
\textbf{Set.} &
\textbf{candidate set} &
\textbf{Negative rule} &
\textbf{Decision context} \\
\midrule
A & $u\!\to\!v$
  & same-day
  & ungrouped, 1:10 \\

B & $u\!\to\!v$ or $v\!\to\!u$
  & same-day
  & ungrouped, 1:10 \\

C & $u\!\to\!v$ or $v\!\to\!u$
  & observed-future
  & ungrouped, 1:10 \\

D & $u\!\to\!v$ or $v\!\to\!u$
  & observed-future
  & same source/day, $K=5$ \\

E & $u\!\to\!v$ or $v\!\to\!u$
  & same-day
  & same source/day, $K=5$ \\

F & $u\!\to\!v$ or $v\!\to\!u$
  & observed-future
  & matched same-source, $K=5$ \\

G & direct or common neighbor
  & same-day
  & same source/day, $K=5$  \\
\bottomrule
\end{tabular}

\caption{
Summary of the seven experimental settings.
Here, $u\!\to\!v$ denotes a historical interaction from the potential blocker to the candidate target.
Settings A--C evaluate ungrouped classification with ten sampled negatives per positive.
Settings D--F compare one positive target with five same-source negative targets and vary the negative definition or matching rule.
Setting G uses the same grouped evaluation but expands eligibility from directly interacting pairs to direct-or-common-neighbor pairs.
}
\label{tab:settings}
\end{table}

\section{Experimental Design}
\label{sec:experimental-setup}

To systematically study how candidate generation and label definitions influence the learning task, we evaluate seven distinct prediction settings, denoted A--G.
These settings vary along three main dimensions: which user pairs are admitted as candidate targets, how negative examples are defined, and whether candidates are evaluated globally or within groups sharing the same potential blocker and prediction day.
Table~\ref{tab:settings} provides a formal summary of our experimental design.

Settings A--C are ungrouped classification tasks.
Setting A considers only pairs in which the potential blocker recently
interacted with the target, while B also admits interaction in the
opposite direction.
Setting C keeps the same bidirectional candidate set as B but
uses the stricter observed-future negative definition.

Settings D and E formulate the task as same-source ranking:
for each future blocker and prediction day, the positive target is
compared with five negative targets associated with the same source.
D uses observed-future negatives, whereas E uses same-day negatives.
Settings F and G retain this grouped decision context but replace
random negative sampling with targeted candidate-selection rules:
\begin{itemize}
    \item \textbf{Setting F (Matched same-source):} We select negative examples that are more similar to the positive target, making the discrimination task more challenging.
We first restrict eligible negatives to the same interaction pattern as the positive pair: $uv$-only, $vu$-only, or reciprocal.
Each eligible target is represented by its historical activity features, log-transformed and standardized. For each positive target, we rank negative targets by Euclidean distance and retain the five nearest cases.
    
    \item \textbf{Setting G (Direct-or-common-neighbor):} We expand the candidate set to include pairs connected either directly or through one common neighbor.
    Let $E_t$ denote the undirected interaction graph constructed from follows, likes, reposts, and replies observed during $[t-W,t-1]$, with $W=\num{7}$.
    Interaction direction and type are ignored when constructing this candidate graph, but remain available as model features.
    A pair $(u,v)$ is eligible when
    $$
    (u,v)\in E_t
    \;\vee\;
    \exists z:
    (u,z)\in E_t
    \wedge
    (z,v)\in E_t
    $$
    that is, $v$ is eligible if they interacted with $u$, or if some user third $z$ interacted with both.
    Since highly connected users can make otherwise unrelated pairs appear structurally close, we exclude the top \num{1}\% of nodes by degree from acting as intermediate nodes $z$.
    These users may still appear as the source $u$ or candidate target $v$.
\end{itemize}

\paragraph{Comparability across settings.}
The settings support targeted comparisons of several experimental choices, although not every cross-setting difference can be attributed to a single factor.
Settings B and C compare same-day and observed-future negative definitions within the same bidirectional candidate set.
Settings D and E examine the same distinction within same-source ranking groups.
Settings D and F assess the effect of replacing random same-source negatives with matched negatives, while Setting G evaluates the effect of expanding the candidate set beyond direct interaction to local network proximity.

\paragraph{Temporal splits.}

We divide the observation period from June~\num{1} to August~\num{28}, \num{2024}, into chronological training, validation, and test intervals.

\begin{itemize}
    \item \textbf{Settings A--B:}
    June~\num{8}--July~\num{30} for training, July~\num{31}--August~\num{14} for validation, and August~\num{15}--\num{28} for testing.

    \item \textbf{Settings C--G:}
    June~\num{8}--July~\num{25} for training, July~\num{26}--August~\num{8} for validation, and August~\num{9}--\num{21} for testing.
\end{itemize}

For Settings C, D, and F, an observed-future negative must remain unblocked through the end of the available data.
We therefore exclude the final \num{7} prediction days so that every retained instance has at least \num{7} days of observable follow-up.
Settings E and G use the same shortened period so that all source-specific ranking settings are evaluated over the same dates.

Settings D, E, and F contain the same positive source-day events and the same number of negatives per positive.
They therefore have identical row counts: \num{447738} training instances, \num{107286} validation instances, and \num{119142} test instances.
Their negative targets differ, however, so the examples are not fully paired across settings.
Setting G contains \num{2197830} training instances, \num{544572} validation instances, and \num{799518} test instances because its candidate rule admits a broader set of user pairs.

\subsection{Model evaluation and interpretation}
\label{sec:model-selection}

Each model assigns a score
$
\hat{p}_{u,v,t}
=
P\!\left(
y_{u,v,t}=1
\mid
\mathbf{x}_{u,v,t}
\right)
$
to every admitted candidate.
The score is conditional on the candidate-generation and negative-sampling procedure used by that setting.
It should therefore not be interpreted as a calibrated probability over all possible user pairs on the platform.

\paragraph{Model selection.}
We first compare several model families on Setting A to identify a strong and computationally practical model to use throughout the remaining experiments.
The comparison includes proximity heuristics, node embeddings, a custom graph neural network, and tabular classifiers.
Table~\ref{tab:model-screening} reports their test-set performance and fitting time.

Simple proximity heuristics, including Common Neighbors, Jaccard similarity, and Salton similarity, achieve at most \num{0.58} ROC-AUC and \num{0.12} AP.
A baseline that ranks candidates only by the target's recent received-block count performs substantially better, despite ignoring the potential blocker.
We refer to this model as the target block-count baseline.

\begin{table}[t]
    \centering
    \small
    \setlength{\tabcolsep}{3pt}
    \begin{tabular}{lccc}
        \toprule
        \textbf{Model} & \textbf{ROC-AUC} & \textbf{AP} & \textbf{Fit time (s)} \\
        \midrule
        Target block-count baseline
            & \num{0.736} & \num{0.389} & $<\num{1}$ \\
        Logistic regression
            & \num{0.875} & \num{0.521} & \num{1244} \\
        Decision tree
            & \num{0.875} & \num{0.554} & \num{132} \\
        MLP
            & \num{0.897} & \num{0.620} & \num{330} \\
        LightGBM
            & \num{0.932} & \num{0.693} & \num{198} \\
        LightGBM + graph embeddings
            & \num{0.931} & \num{0.698} & -- \\
        Multi-relational GNN
            & \num{0.919} & \num{0.648} & \num{1624} \\
        \bottomrule
    \end{tabular}
    \caption{
Model-group screening on the Setting A test split.
Fit times correspond to the implementations used in the screening and
include model-specific preprocessing within each training pipeline,
while excluding upstream feature construction.
}
    \label{tab:model-screening}
\end{table}

The graph neural network is a custom multi-relational model based on \citet{schlichtkrull2018modeling}.
It applies a separate transformation to each interaction type, followed by two layers of degree-normalized message passing.
A feed-forward decoder scores each source--target pair from the resulting node representations.

LightGBM achieves the highest ROC-AUC and nearly the highest AP in the screening experiment.
Its AP differs from the graph-embedding variant by only \num{0.005}, while its fitting time is approximately eight times lower than that of the custom GNN.
It also avoids the graph-snapshot preprocessing required by the embedding and GNN models.
We therefore use LightGBM as the prediction model of choice, and train a separate instance for each prediction setting.
This choice is also consistent with prior evidence that gradient-boosted trees perform strongly on medium-sized tabular datasets \cite{grinsztajn2022tree}.

\paragraph{Evaluation metrics.}

We report ROC-AUC, average precision, precision, recall, and F1.
Because block events remain sparse within the constructed datasets, we use average precision as the main threshold-independent classification metric.
For threshold-dependent metrics, the classification threshold is selected by maximizing F1 on the validation set and is then applied unchanged to the test set.

For Settings D--G, candidates are grouped by source and prediction day.
Each group contains one positive target and ${K=5}$ negative targets.
We report tie-aware Hits@1, Hits@3, and mean reciprocal rank for these settings.
Under random ranking, the expected Hits@1 baseline is
$
\frac{1}{K+1}
=
\frac{1}{6}
\approx
\num{0.167}.
$
These ranking metrics remain conditional on the positive target being admitted by the corresponding candidate rule and on comparison with the sampled negatives.

\paragraph{Uncertainty and model interpretation.}

We estimate uncertainty on test-set performance using \num{1000} cluster-bootstrap resamples, without refitting the models.
For ungrouped Settings A--C, prediction days are resampled with replacement.
For grouped Settings D--G, complete source-day ranking groups are resampled with replacement to preserve the within-group structure.
For each resample, performance metrics are recomputed from the corresponding test-set predictions, and we report the standard deviation across bootstrap estimates.

We examine model behavior using gain-based feature importance, mean absolute SHAP values \cite{lundberg2017unified}, and feature-group ablations.
The ablations measure the change in performance after removing pair-, source-, target-, graph-, or content-level features.
All interpretations are relative to the candidate and sampling protocol of the corresponding setting.

Code and reproducibility details are provided
in the anonymized Code and Data Supplement.

\section{Results}

\subsection{Exploratory analysis of pre-block interactions}
\label{sec:direct-interaction}

\begin{figure}[t]
    \centering
    \includegraphics[width=0.65\linewidth]
    {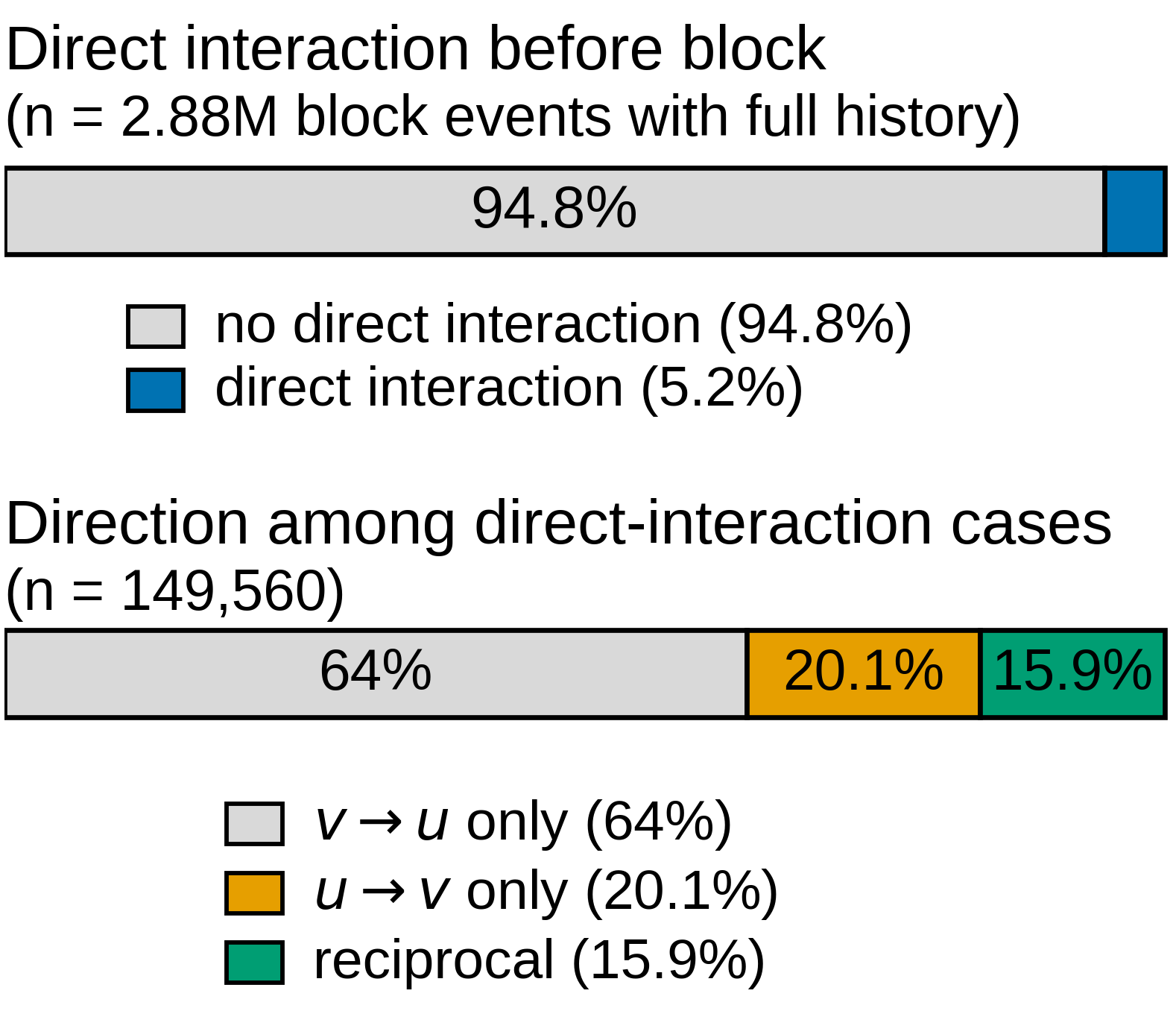}
    \caption{
    Only $5.2\%$ of future blocks are preceded by direct interaction.
    Among these, interaction from the eventual target toward the
    eventual blocker is the most frequent pattern.
    }
    \label{fig:direct-interaction-before-blocking}
\end{figure}

We first examine whether observed block events are preceded by recent interaction between the future blocker and target. 
Among 2,878,871 first observed blocks with a complete seven-day history, only 149,560 ($5.2\%$) involve direct interaction between the two users during the historical window (Figure~\ref{fig:direct-interaction-before-blocking}). 
The remaining $94.8\%$ occur between pairs with no recent direct interaction.

When direct interaction is present, it is usually asymmetric. 
Among these cases, $64.0\%$ involve only interaction from the future target to the future blocker ($v\!\to\!u$), $20.1\%$ involve only interaction from the future blocker to the target ($u\!\to\!v$), and $15.9\%$ involve interaction in both directions. 
Thus, the most common pre-block pattern is interaction initiated by the account that is subsequently blocked.

Block pairs without recent direct interaction may still be connected through the local interaction network. 
Applying the top-$1\%$ intermediary-degree filter used in Setting~G, $65.3\%$ of these pairs are connected by a path of length at most four. 
When the analysis is restricted to pairs whose endpoints are both present in the historical interaction graph, this proportion rises to $85.7\%$, compared with $76.4\%$ for same-source, degree-matched non-block pairs (Table~\ref{tab:no-direct-interaction-distance}).

\begin{table}[t]
\centering
\small
\begin{tabular}{lcc}
\toprule
\textbf{Distance} & \textbf{Blocks} & \textbf{Matched non-blocks} \\
\midrule
2 & 23.06 & 12.03 \\
3 & 42.20 & 35.03 \\
4 & 20.39 & 29.44 \\
$>4$ or disconnected & 14.35 & 23.50 \\
\bottomrule
\end{tabular}
\caption{
Shortest-path distance distribution (\%) for block pairs with no recent
direct interaction and same-source, degree-matched non-block pairs.
}
\label{tab:no-direct-interaction-distance}
\end{table}

\begin{table}[t]
\centering
\small
\setlength{\tabcolsep}{3pt}
\begin{tabular}{lcc}
\toprule
\textbf{Candidate rule} & \textbf{Coverage} & \textbf{Common Neighbor gain} \\
\midrule
Direct dyadic & 5.16\% & -- \\
Direct or common neighbor & 40.96\% & 35.80 pp \\
Setting G & 22.01\% & 16.85 pp \\
\bottomrule
\end{tabular}
\caption{
Platform-wide candidate coverage before negative sampling.
Common-neighbor gain is the additional coverage contributed by
common-neighbor-only pairs.
Setting~G excludes the top $1\%$ of nodes by degree as intermediaries.
(approximately $22.6\%$).
}
\label{tab:g-coverage}
\end{table}

We complement this pair-level analysis with a platform-wide audit of the candidate-generation rule used in Setting~G before negative sampling (Table~\ref{tab:g-coverage}). 
A rule based only on recent direct interaction admits approximately $5.2\%$ of observed block events. 
Expanding the candidate set to include common-neighbor pairs increases coverage to approximately $41.0\%$ without hub filtering and to approximately $22.0\%$ when the top $1\%$ of nodes by degree are excluded as intermediaries, as in Setting~G. 
Under this filtered rule, common-neighbor-only pairs contribute an additional $16.9$ percentage points of coverage.

\subsection{Observed-future filtering has limited impact}
\label{sec:predicting-blocks}

\begin{figure}[t]
    \centering
    \includegraphics[width=0.82\columnwidth]{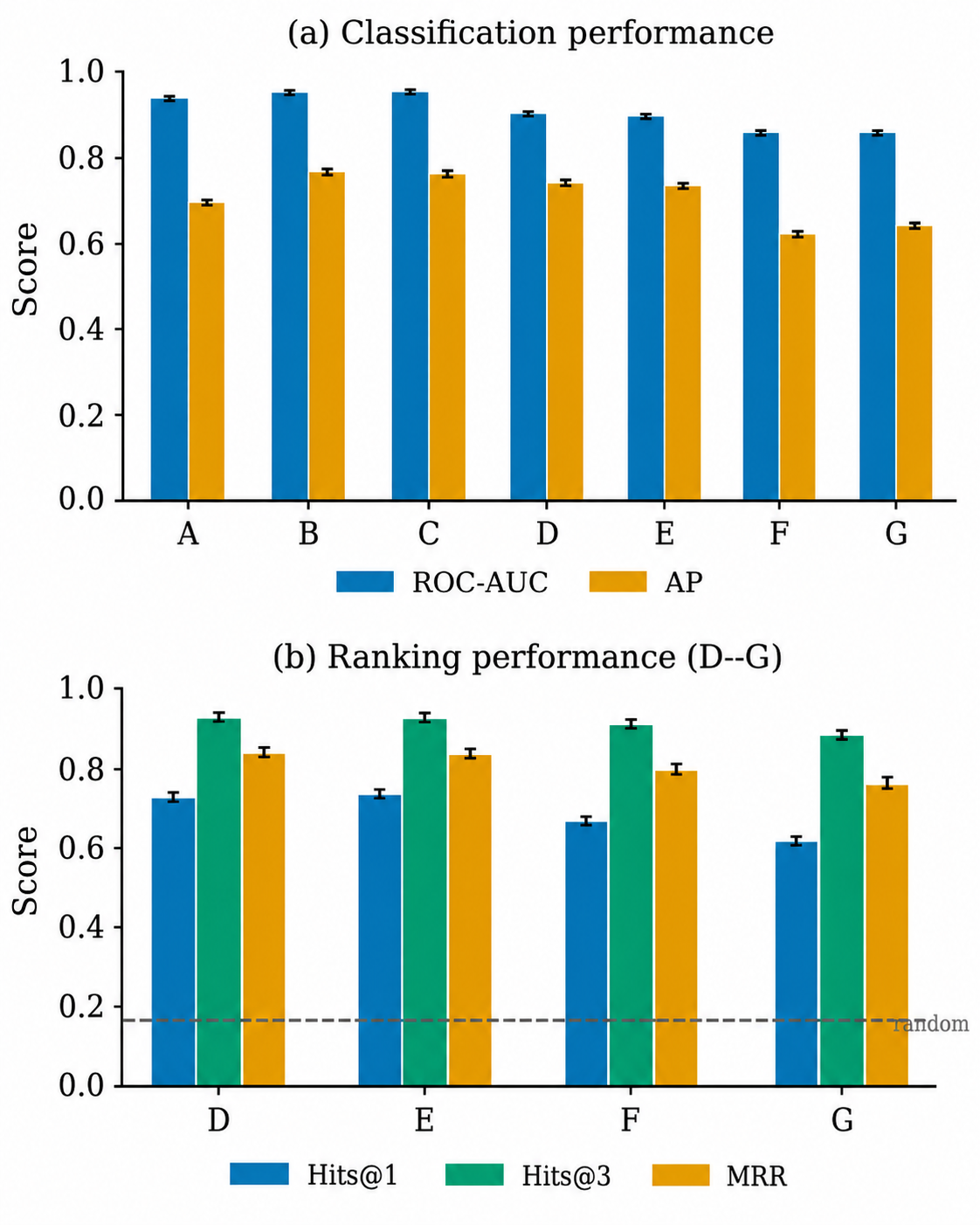}
    \caption{
    Predictive performance across experimental settings.
    Panel (a) reports ROC-AUC and average precision for Settings A--G.
    Panel (b) reports Hits@1, Hits@3, and MRR for the same-source ranking
    Settings D--G; the dashed line indicates the random Hits@1 baseline
    of $1/6$.
    Error bars show cluster-bootstrap standard deviations over 1,000 resamples.
    }
    \label{fig:performance-summary}
\end{figure}

Figure~\ref{fig:performance-summary} summarizes predictive performance across all seven settings; complete numerical results are reported in the Code and Data Supplement.

Settings B and C yield nearly identical average precision. Because their primary test periods differ, we also reran the comparison over their shared August~15--21 window. 
Average precision was $0.782$ for Setting~B and $0.781$ for Setting~C. 

Settings D and E, which use the same evaluation dates, candidate set, grouping structure, and number of examples, likewise produce similar classification and ranking performance. 
These comparisons suggest that, under the sampling protocols considered here, replacing same-day negatives with observed-future negatives has little effect on measured performance. 
However, the negative targets in Settings D and E are sampled separately, so the results are not based on fully paired candidate groups.

Performance decreases when the prediction task becomes more challenging. 
Setting~F compares each positive target with behaviorally similar negative targets, whereas Setting~G expands the candidate set to include common-neighbor pairs. 
Hits@1 decreases from $0.724$ in Setting~D and $0.723$ in Setting~E to $0.657$ in Setting~F and $0.613$ in Setting~G. 
All four values remain substantially above the random-ranking baseline of $0.167$ (Figure~\ref{fig:performance-summary}).
In Setting~G, the model ranks the true target first in $61.3\%$ of groups containing one positive target and five same-source negative targets. 
This result measures ranking performance only after the true target has been admitted to the candidate set. 
The platform-wide observability audit indicates that the Setting~G candidate-generation rule admits approximately $22\%$ of observed block events. 
Thus, the ranking result should be interpreted as conditional performance rather than platform-wide recall.

\subsection{Experimental design changes which features matter}
\label{sec:interpretability}

\begin{figure}[t]
    \centering
    \includegraphics[width=\linewidth]{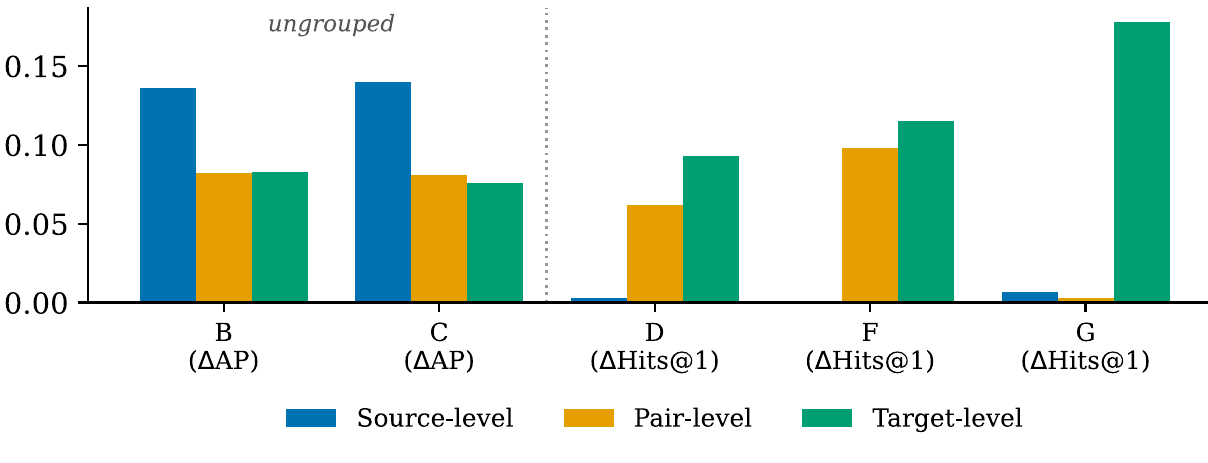}
    \caption{
    Change in test performance after removing one feature group.
    For Settings B and C, the bars report the decrease in AP;
    for Settings D, F, and G, they report the decrease in Hits@1.
    Larger values indicate a larger performance decrease after removing
    the corresponding feature group.
    Source-level features have the largest effect in ungrouped classification,
    while target-level features have the largest effect in same-source ranking.
    }
    \label{fig:ablation-summary}
\end{figure}

The feature-group ablations reveal the clearest differences across settings (Figure~\ref{fig:ablation-summary}). 
In the ungrouped classification Settings~B and C, removing source-level features causes the largest decrease in AP, by $0.136$ and $0.140$, respectively. In the same-source ranking Settings~D, F, and G, however, all candidates within a group share the same source user. 
Source-level features therefore vary little within groups and contribute less to ranking targets. 
Instead, removing target-level features produces the largest decrease in Hits@1: $0.093$ in Setting~D, $0.115$ in Setting~F, and $0.178$ in Setting~G. Pair-level features have a larger effect in Settings~D and F than in Setting~G, where many candidates are connected only through a common neighbor and have no recent direct interaction. 
These results show that feature importance depends not only on the underlying behavior but also on how candidate targets are selected, grouped, and compared.

Across Settings~B--G, the number of recent blocks received by the target is the strongest individual predictor according to both gain-based feature importance and mean absolute SHAP. 
Used as the only predictor, this feature achieves ROC-AUC values of $0.722$--$0.805$ and AP values of $0.354$--$0.510$. 
Removing it from models restricted to the ten highest-ranked features reduces AP by $5.9$--$14.5\%$. Recent received-block count therefore contains substantial predictive information, but the mechanism underlying this association is unclear. 
It may reflect harmful conduct, greater visibility, controversy, polarization, coordinated exclusion, or several of these factors simultaneously.
We also find considerable redundancy among the available predictors: approximately $8$--$9\%$ of the features recover $97$--$99\%$ of the achievable AP across settings.

Directional interaction provides a second example of how experimental design affects the observed associations. In Setting~D, candidates with target-to-source interaction---particularly cases in which the target follows the source---have a positive rate of $54.0\%$, well above the group base rate of $16.67\%$. By contrast, reciprocal interaction is associated with a positive rate close to the base rate. In Setting~F, where negative targets are selected to be more similar to the positive target, the positive rate associated with target-to-source interaction falls to $33.8\%$. This reduction suggests that part of the association observed in Setting~D may result from differences between the positive targets and randomly sampled negatives, rather than from the directional interaction pattern alone. Because the analysis is observational, these findings should not be interpreted as evidence that target-to-source interaction causes blocking.

A different pattern emerges in Setting~G, which admits candidates connected through common neighbors as well as direct interactions. Under this broader candidate set, local graph structure becomes more informative for distinguishing among targets. Adamic--Adar similarity ranks among the strongest features by mean absolute SHAP and, when used alone, achieves a ROC-AUC of $0.622$ and an AP of $0.279$, compared with a positive base rate of $16.7\%$. 
These results further show that the signals associated with predictive performance depend on the candidate-generation rule.

\section{Discussion and Conclusion}

We formulate directed block prediction on Bluesky as a candidate-conditioned temporal task and examine how candidate set construction and evaluation protocol influence measured performance and feature relevance.
Candidate selection based on recent direct interaction only capture a small share of observed block events, while expanding the candidate set through common neighbors substantially improves coverage. 
Conditional on the true target---the negative case---being admitted, the models rank it above sampled alternatives substantially more often than random. 
The ablation results further show that the apparent importance of source\nobreakdash-, target\nobreakdash-, and pair-level features changes across evaluation settings. 
These findings highlight that candidate coverage should be evaluated separately from classification and ranking performance, which is conditional to the generated candidate sets.
These results have implications for both predictive evaluation and moderation research. 
Candidate generation determines which accounts can be considered and therefore places an upper bound on the block events that a prediction framework can capture. 
A model may perform well among admitted candidates while remaining unable to identify most future block targets. 
Model performance and feature interpretation should therefore be reported relative to the candidate-generation rule, grouping structure, and negative-sampling procedure used in each setting.

The strong association between the number of recent blocks received by a target and model performance also deserves attention. 
This feature may reflect harmful behavior, but it may also capture visibility, controversy, polarization, coordinated blocking, or a combination of these factors. 
Any user-facing application should therefore not treat prior blocks received as direct evidence of harmful conduct, and should remain opt-in, preserve user control, and complement rather than replace platform-level moderation and enforcement~\cite{jhaver2025individualcontrol,jhaver2023personalizing}.

Our work is not without limitations. 
The results are specific to the observed Bluesky data and experimental settings.
Candidate coverage and ranking performance use different denominators, so the latter should not be interpreted as platform-wide recall. 
The observational data also do not establish why users block or whether predictive suggestions would improve safety. 
Moreover, directed block prediction is dual-use and could support block evasion or harassment.

Future work should develop candidate-generation rules that improve coverage without producing excessively large or irrelevant candidate sets. 
It should also compare more sophisticated models and test robustness across longer periods and user groups, evaluate effects on users' safety and control, and study the causal effects of blocking on subsequent interactions, exposure, and user behavior. 

From an impact perspective, this work highlights that the social consequences of a predictive system depend not only on model performance, but also on which users the system considers for the prediction, which in turn influences the behavioral signals that the model relies on. 
Treating candidate generation, user control, and risks of unfair exclusion as part of the core evaluation is therefore essential when studying AI-assisted moderation.

\bibliography{references}

\end{document}